\documentclass{elsart}
\usepackage{amsmath,amsfonts,amssymb,color,graphicx}
\begin{document}

\begin{frontmatter}

\title{Rapid Spectral Variability during the 2003 Outburst of
V4641 Sgr (= SAX J1819.3-2525)\thanksref{grant}}
\author[yale]{Dipankar Maitra},
\ead{maitra@astro.yale.edu}
\author[yale]{Charles Bailyn}
\address[yale]{Yale University, New Haven, CT 06511, USA}
\thanks[grant]{This work was supported by the National Science Foundation grant
AST 0407063.}

\begin{abstract}
The black hole candidate V4641 Sgr (= SAX J1819.3-2525) went through a brief
outburst during 2003 Aug 01 to Aug 08. During the outburst, activity was noted
in optical, radio as well as X-rays. Here we present results of Spectral and 
Temporal analysis of a pointed {\em Rossi X-ray Timing Explorer} (RXTE) 
observation of the source during the outburst. During this pointing we observed 
flaring activities with associated X-ray luminosity variations over factors of 
5 or more in timescales of few tens of seconds. The observed flares during
are intrinsically different in their spectral and temporal properties. We see 
evidence of variable column density during one of the flares. The spectral and 
temporal analyses of the data suggest occasional outflow/mass ejection.
\end{abstract}

\end{frontmatter}

%\section{Introduction}
The compact binary system V4641 Sgr (= SAX J1819.3-2525) has been known to 
exhibit short, irregular outbursts, with highly variable lightcurve profile. 
The rapid time variability \cite{r2002,w2000} in the lightcurve as well as 
observation of moving radio structure \cite{h2000} during an outburst in 1999 
led to its classification as a microblazar \cite{mr1994}.

\begin{figure*}[ht]
\centering
\includegraphics[width=0.5\textwidth]{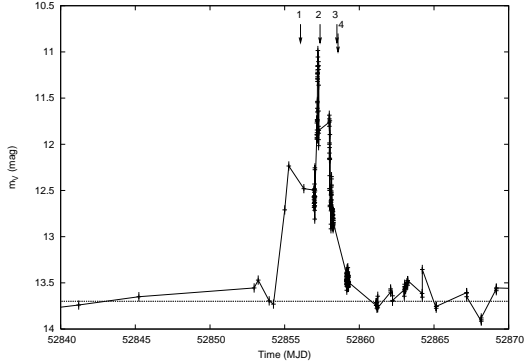}
\caption{
Optical V band lightcurve showing the 2003 outburst of V4641 Sgr.
Four vertical lines on the top labelled 1-4 are the times when significant
X-ray activity was observed by RXTE. The horizontal dotted line represents
the mean quiescent brightness of 13.7 magnitude. The data were taken by the CTIO
1.3m telescope operated by the SMARTS Consortium.
\label{fig:V-band}}
\end{figure*}

During early Aug 2003, signs of enhanced activity were reported by the 
VSNET\footnote{http://ooruri.kusastro.kyoto-u.ac.jp/mailman/listinfo/vsnet-alert} group and shortly thereafter by the {\em Small and Moderate Aperture Research
Telescope System}
consortium telescopes at Cerro Tololo Inter-American Observatory (CTIO) in
Chile \cite{atel170}. In Fig.~\ref{fig:V-band} we show the V band lightcurve 
of V4641 Sgr during this outburst, as observed by the SMARTS 1.3m telescope. 
We used the CTIO observations to trigger the series of 
target-of-opportunity (ToO) X-ray observations. During four pointed RXTE 
observations we saw activity from the source.  The observed X-ray spectrum 
during all these 4 pointings were hard in nature, showing large contribution 
of hard X-rays. Strong Fe K$\alpha$ fluorescent emisssion line near 6.5 keV 
was detected with large equivalent widths in the range of 700 - 1000eV.

\begin{figure*}[ht]
\centering
\includegraphics[width=0.7\textwidth]{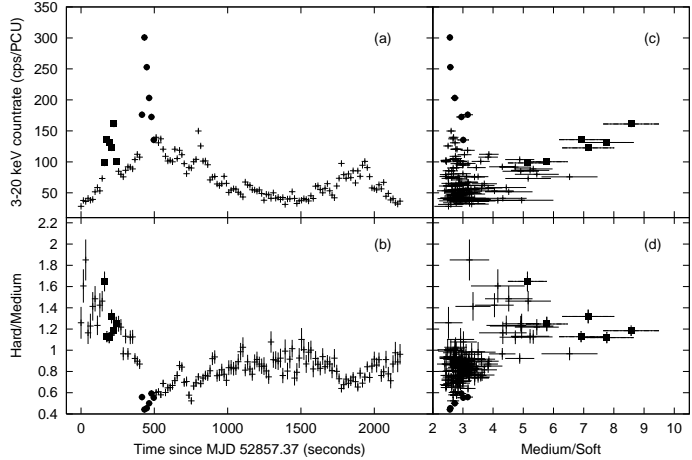}
\caption{
RXTE/PCA observations of V4641 starting MJD 52857.37 are shown. 
The data during the hard flare are shown by $\blacksquare$, those during soft 
flare are shown by $\bullet$ and the rest of the dwell are shown by + symbols. 
Panel (a): The 3-20 keV lightcurve showing the rapid variability of this source 
that sets it apart from typical X-ray binaries. 
Panel (b): Variation of hardness-ratio with time. Hard band is 10.3-20.4 keV, 
medium band is 5.3-10.3 keV, soft band is 2.0-5.3 keV.
Panel (c): Color-Luminosity plot showing the different regions occupied by the
soft and hard flares.
Panel (d): Color-color plot, also shows the while the soft flare occupies the 
most soft spectral state (lower-left corner), the hard flare essentially 
occupies the extreme hard state in the upper-right corner of the color-color
diagram.
\label{fig:lc}}
\end{figure*}

In this article we concentrate, in particular, on second dwell where we 
observed two flaring activities with flare timescales of tens of seconds. In
panel (a), (b), (c) and (d) of Fig.~\ref{fig:lc} we show the evolution of the 
source lightcurve, hardness ratio, color-magnitude diagram and color-color 
plane respectively during dwell (2). The distinctively opposite characters of 
the two 
flares are evident. While the first flare at $\sim250$ seconds is essentially 
hard, the second flare during $\sim450$ second is predominantly soft. 

\begin{figure*}[ht]
\centering
\includegraphics[width=0.39\textwidth, angle=-90]{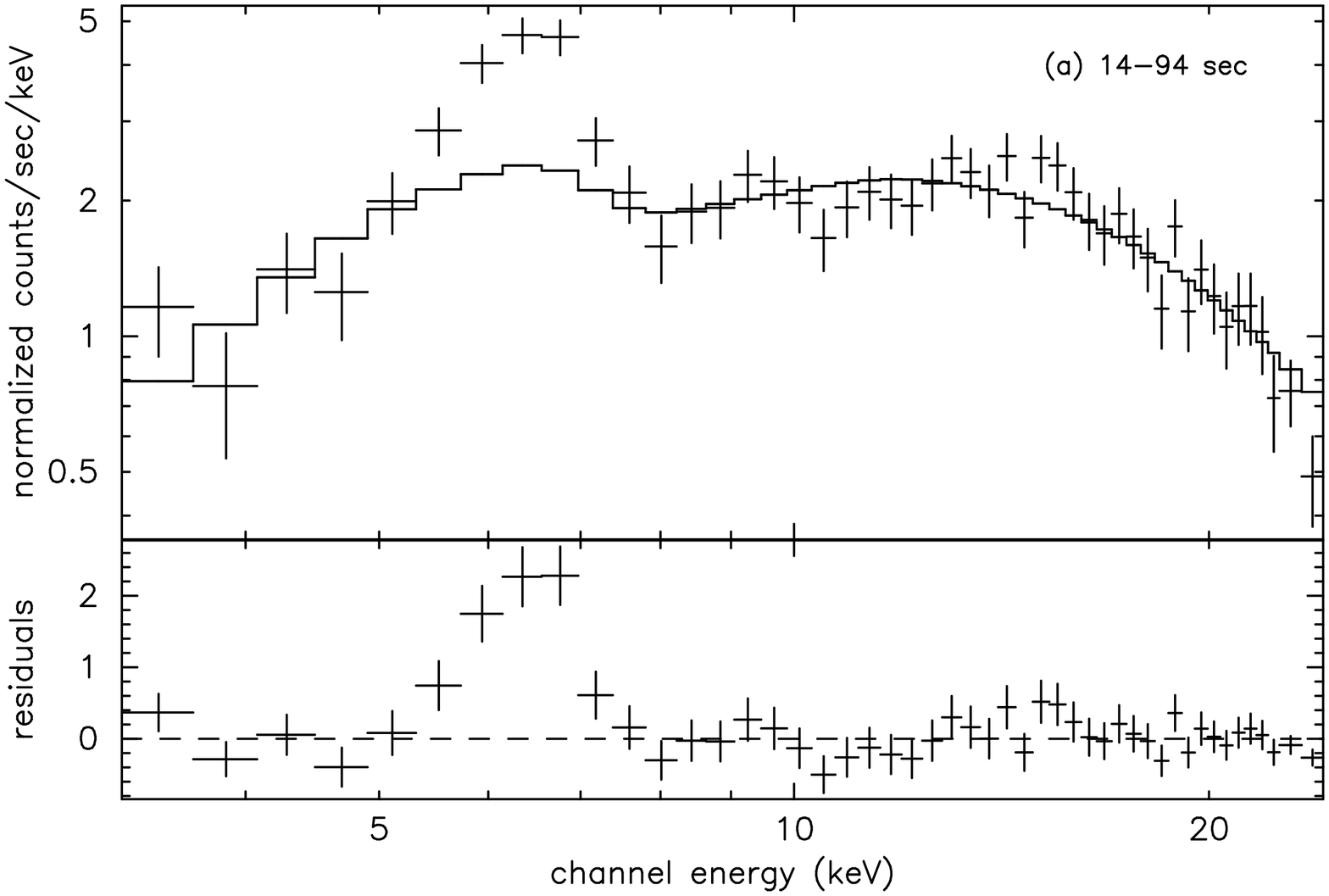}
\includegraphics[width=0.39\textwidth, angle=-90]{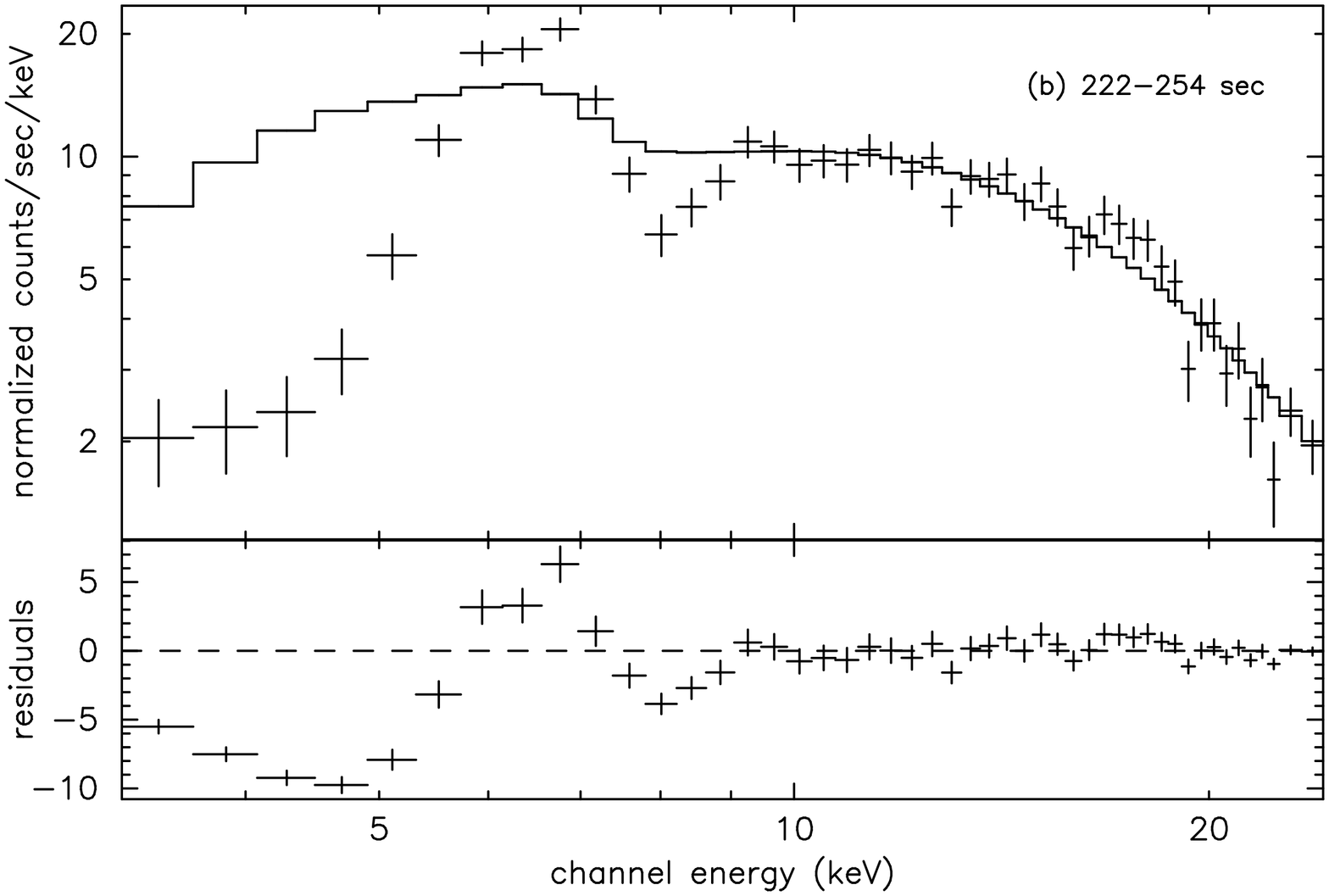}
\caption{Variation of $n_H$ during the hard flare in dwell (2). Panel (a):
The energy spectrum during 14-94 seconds. Panel (b): The spectrum during
222-254 seconds when the source was going through a hard flare. The solid
histogram in both panels is a fit to the 10.0-25.0 keV spectrum with
$n_H=2.3\times10^{21}$ $atoms/cm^2$. Note that the
extrapolated fit matches the continuum at lowest energies for panel (a)
whereas it grossly overestimates the counts during the flare as shown in
panel (b).\label{fig:nh_spec}}
\end{figure*}

\begin{figure*}[ht]
\centering
\includegraphics[width=0.5\textwidth, angle=0]{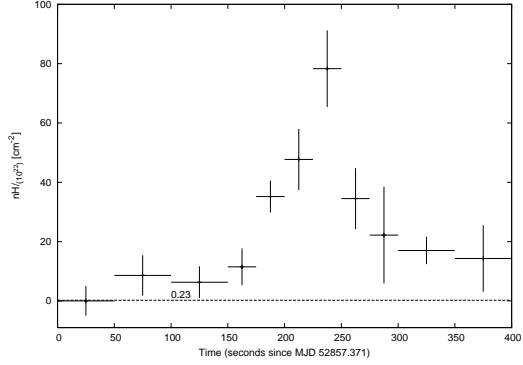}
\caption{Variation of the fit parameter $n_H$ during the hard flare. The
dashed horizantal line represents the standard adopted value of $n_H$ for
this source.
\label{fig:nh_vary}}
\end{figure*}

Since the 
Iron line dominates the low energy spectrum, estimating the continuum becomes 
difficult. We modelled the X-ray spectrum in the 10-25 keV region using the 
{\em pexrav} model \cite{mz1995} which calculates an exponentially cut
off power law spectrum reflected from neutral material. The spectrum during the
first 14-94 seconds is well fit by this model and a warm absorber with its 
column density ($n_H$) set to the standard value of 
$2.3\times10^{21}$ atoms/cm$^{-2}$ for this source \cite{dl1990}.
The fit model, when extended to softer energies, fails to account for the 
Fe $K\alpha$ line near 6.5 keV, but matches the continuum at the softest bands
as shown in the left panel of Fig.~\ref{fig:nh_spec}. However, during the 
hard 
flare, the fits to high energy regimes of the spectrum ($>10$ keV) largely 
overestimates the counts in softer energies as shown in the right panel of 
Fig.~\ref{fig:nh_spec}. In fact, we were unable to find reasonable fits with 
any physically motivated models with $n_H$ fixed to its standard value. 
Allowing $n_H$ results in much better fits. In Fig.~\ref{fig:nh_vary} we show 
the variation of the fit parameter $n_H$ during the hard flare.

%\clearpage
{\bf Conclusions}
\begin{itemize}
\item
V4641 Sgr went through a short lived, hard flaring outburst between 2003 Aug
01 through Aug 08, during which X-ray fluctuations by a factor of ~10 on
timescales of tens of seconds were seen.
\item
Compton reflection features like strong Fe $K\alpha$ line near 6.5 keV and 
curvature in spectrum at high energies were present in all observations.
\item
Spectral nature of some flares separated in time only by few minutes were 
intrinsically different.
\item
The thick warm envelope/outflow scenario \cite{r2002,ts2005} can
explain the variation of column density during the second pointing although 
other scenarios like variation of partial covering fractions cannot be ruled 
out.
\end{itemize}

\end{document}